# The Role of Percolation and Sheet Dynamics during Heat Conduction in poly-dispersed Graphene Nanofluids

Purbarun Dhar[1], Soujit Sengupta[2], Saikat Chakraborty[1], Arvind Pattamatta[1] and Sarit K. Das[1,*]

[1]Department of Mechanical Engineering, Indian Institute of Technology, Madras, Chennai – 600 036, India

[2]Department of Chemistry, Indian Institute of Technology, Madras, Chennai – 600 036, India

**Abstract**

A thermal transport mechanism leading to the enhanced thermal conductivity of Graphene nanofluids has been proposed. The Graphene sheet size is postulated to be the key to the underlying mechanism. Based on a critical sheet size derived from Stokes-Einstein equation for the poly-dispersed nanofluid, sheet percolation and Brownian motion assisted sheet collisions are used to explain the heat conduction. A collision dependant dynamic conductivity considering Debye approximated volumetric specific heat due to phonon transport in Graphene has been incorporated. The model has been found to be in good agreement with experimental data.

Graphene, the two dimensional material, with its typically high thermal and electrical conductivities[1], has become a major focus for the scientific community over the last decade, and has also found its way into nanofluid (dilute suspensions of nano-particles in conventional coolant liquids) research. In the present era of miniaturization and/or enhanced capacity, devices all around the globe pose severe cooling challenges due to generation of unprecedentedly high heat flux. Conventional coolants prove ineffective in such cases and might lead to device failures due to high thermal loads. This is where nanofluids have edged in as potential future coolants. Graphene nanofluids may also emerge as potential smart-fluids with enhanced electrical conductivities[2], the starting material for manufacturing Graphene thin films and as carrier agents for targeted drug delivery. In this Letter, the physics behind heat conduction in Graphene nanofluids (GNF) has been investigated. For the first time, the GNF has been treated as a poly-dispersed system (as opposed to the conventional approach of considering a mono-dispersed system of an average particle size) and an analytical model to predict its thermal conductivity has been proposed.

A plethora of nano-particles have been used by the scientific community to study their effects on liquids, ranging from metallic[3,4], metal oxide[3,5,6] and carbon based[2,7-9] nanoparticles. Among these, the carbon based nano-particles provide much higher enhancements at much lower concentrations[7]. Initially, CNTs were thought to exhibit the best results due to their very high thermal conductivity ($k_{CNT} > 3000$ Wm$^{-1}$K$^{-1}$ [10]), but recent experiments [2,11,12] on GNFs show much higher enhancements than CNT, for similar volume concentrations. This could be initially attributed to the much higher thermal conductivity of Graphene (~ 5000 Wm$^{-1}$K$^{-1}$ [13]). However, the experimental data[11] shows a sharp rise in the effective conductivity of the GNF with temperature, a phenomenon absent in CNT nanofluids and also grossly different from the trend shown by metal and metal-oxide nanofluids, bringing to the forefront a possible dynamic mechanism of enhancement in GNFs. Thus, attributing the high thermal conductivity of Graphene as the sole reason for such enhancement would not be justified. The phenomenon of temperature dependent thermal conductivity, first reported by Das et.al.[14], remains one of the most significant discoveries in nanofluid research. Over the years, numerous models have been proposed in literatures to explain heat conduction in nanofluids, based upon various mechanisms of heat transport, viz. percolation theory for nanoparticles [15-17], Brownian motion induced thermal transport [18-20], micro-convection due to random motion of nanoparticles[21,22], phenomenon of liquid layering [16,23] etc. However, it has been observed that none of these models are able to predict the temperature dependent thermal conductivity of GNFs accurately. Temperature is found to have a much stronger influence on the thermal conductivity of the GNF than predicted by most of these models. Furthermore, the conventional approach of viewing nanofluids as a mono-dispersed system, with an average particle size, leads to significant loss of accuracy in predicting its thermal properties. Hence, a model (considering the GNF as a poly-dispersed system) to predict the effective thermal conductivity of GNFs, simultaneously providing insight into the underlying physical mechanisms involved behind its strong temperature dependence is the need of the hour.

The proposed model has been evolved considering an elemental analytical domain or cell within the GNF. The thermal transport in GNFs is highly dependent on the sheet-size distribution of the Graphene sample, due to the fact that Graphene samples exhibit a wide range variation in sheet sizes which leads to a poly-dispersed nanofluid. Two physical phenomenon, (elaborated at a later stage) : (i) Sheet percolation and (ii) Temperature dependant sheet dynamics, are the major agents governing the heat conduction within this domain. An illustration of the analytical domain has been provided in Fig.(1).

**[FIGURE 1]**

In order to determine the fraction of the poly-dispersed nano-sheet population contributing towards each of the two mechanisms, it has been proposed that a critical size order exists for such nanosheets. It is such that the sheets with characteristic sizes (sheet face length) below the critical size shall be strongly affected by the thermal motion of the fluid molecules and will thus enhance the thermal conductivity of the system due to dynamicity. The sheets larger than the critical size however, are large enough to show more resilience to the thermal motion of the fluid molecules than the smaller sheets. Furthermore, their sizes allow them to form stable percolation networks and enhance the thermal conductivity of the system through the conductive percolation chains. It is proposed that the critical size order can be estimated by equating the unidirectional Brownian velocity ($U_B$) of the sheet, as given by Stokes-Einstein's formula in Eq. (1),

$$U_B = \frac{2k_B T}{3\pi \mu L_g^2} \quad (1)$$

to the sheet settling velocity (Eq. (2)) (also obtained from Stokes-Einstein's formula).

$$v_s = \frac{2L_g^2 |\rho_p - \rho_m| g}{36\mu} \quad (2)$$

$k_B$ is the Boltzmann constant, T is the absolute temperature of the system, μ is the dynamic viscosity of the base fluid, $L_g$ is the effective face size of the nanosheets, $v_s$ is the settling velocity, $\rho_p$ is the density of graphene and $\rho_m$ is the density of base fluid. An assumption, that Stokes-Einstein's law is equally valid for nanosheets as it is for nano-spheres, has been made in this approach. These equations can be used to determine the critical sheet face length value $L_{cr}$, for which $U_B$ and $v_s$ are of equal magnitudes. Sheets of sizes very close to or larger than $L_{cr}$ are considered to constitute the distribution fraction 'α' (0≤ α ≤1). 'α' can be obtained from Dynamic Light Scattering (DLS) or similar analysis of the GNF.

Having established a critical size, the complete description of the analytical model can be elaborated as follows:

(*i*) Sheet Percolation: The GNF is assumed to be wholly consisting of Graphene sheets of face size similar to or larger than the critical size order $L_{cr}$, i.e. α is unity. Furthermore, for simplicity, the face area of the sheets, (whatsoever be the face-shape) is assumed equivalent to $L_g^2$. The present approach is based on percolation theory[15] which uses the concept of thermal transport along highly conductive paths created by the percolation of nanoparticles in the base medium. The present model assumes graphene nanosheets as equivalent solid flat plates, linked to one another in a random fashion. The calculations for the effective percolating length of individual nanosheets, their orientation in 3 dimensional space and determining the dimensions of the analytical domain are based on percolation theory[15]. While evaluating for the total number of parallel Graphene chains within the domain, it is necessary to utilize the effective area of heat conduction for a flat sheet. The total number of parallel graphene chains (M) in the domain is evaluated as:

$$M = \frac{\varphi L_{cell}^3}{(n_g-1)d_g L_g^2 N} \quad (3)$$

where, $\varphi$ is the volume percentage of Graphene loading, $n_g$ is the average number of layers for the Graphene nanosheets, $d_g$ is the inter-sheet distance for the Graphene sample and N is the number of nanosheets in one single percolation chain. The term $(n_g - 1)d_g L_g^2$ denotes the volume of each nanosheet and does not exist for pure single layer graphene.

The present model utilizes a net resistance approach[15] to determine the effective thermal transport due to heat conduction along the percolation chains, conduction through the fluid itself, conduction from the fluid to individual percolating sheet and vice versa. The resistance offered by each Graphene sheet, $R_G$, is determined based on Fourier heat conduction through a flat plate. The interfacial contact resistance between individual sheets and the fluid has been determined based on the effective area of contact between two flat plates with a thin liquid film in between. The interfacial contact resistance is determined as:

$$R_c = \frac{1}{A_{contact}G} = \frac{1}{(n_g-1)d_g L_g G} \quad (4)$$

where, G is the interfacial thermal conductivity between a Graphene sheet and the base fluid. Huxtable et.al.[24] determined the value of G for CNT-water interface to be 12 MWm$^{-2}$K$^{-1}$. Since the in-plane thermal conductivity of Graphene is nearly double that of CNT, the interfacial conductance value at the Graphene-water interface is assumed to be double that of CNT-water interface, and 25 MWm$^{-2}$K$^{-1}$ has been used as the value of G for the present study. The net interfacial contact resistance between two neighboring sheets is effectively $2R_c$. The net heat conductance of the analytical domain is then computed for tri-layer Graphene (with the standard 0.335 nm inter-layer distance[1]) of average sheet face size of 1.5 microns, as obtained from DLS analysis[11]. A computer code has been used to evaluate the model, wherein, random number generators are used to assign random values for the percolating length and the sheet orientations in space. The code is evaluated for values of N in the range of $10^3$ or greater, since higher values ensure a more normalized distribution of the random variables. The variation of thermal conductivity values for the domain due to percolation is found to be within 0.5% for multiple runs of the code. The average value of five such runs is considered as the final thermal conductivity value for the domain, and is represented as $k_{perc}$.

(ii) The temperature dependant sheet dynamics: The GNF is assumed to be composed entirely of nanosheets of an average face size which is smaller than the critical face size, i.e. α is considered to be zero. In essence, the assumption is that the sole mechanism of thermal transport within the whole analytical domain is Brownian motion assisted inter-sheet collisions in addition to effective medium theory (EMT)[20]. In the present study, the thermal conductivity due to sheet dynamics, $k_{sd}$, is theorized to consist of three parts, as:

$$k_{sd} = k_{medium} + k_{EMT} + k_{dynamic} \qquad (5)$$

$k_{EMT}$ is the enhancement in the thermal conductivity of the fluid solely due to the presence of the nanosheets. Based on an analogical treatment of EMT, keeping in mind the structure and special features of Graphene, $k_{EMT}$ is determined as:

$$k_{EMT} = \frac{k_p \varphi d_m}{(1-\varphi)L_g} \qquad (6)$$

where, $k_p$ is the in-plane thermal conductivity of Graphene and $d_m$ the molecular diameter of the fluid. Although EMT holds good for spherical nano-particles, sheet dynamics is appreciably high only for sheet sizes well below the critical size, and so it is assumed that flat sheets behave similar to spherical particles. Hence, particle diameter has been replaced by sheet face size. However, unlike dynamic EMT[20], in the present case, it is postulated that the dynamic conductivity is completely independent of $k_{EMT}$ and exists due to Brownian motion assisted inter-sheet collisions within the fluid domain.

The dynamic conductivity has been theorized to be a function of all the factors that accurately describe the dynamic heat transport behavior of a nano-particle within a fluid domain. It has been proposed that

$$k_{dynamic} = \overline{U_B} \lambda C_v \varphi \theta \qquad (7)$$

where, $\overline{U_B}$ represents the mean Brownian velocity of the nanosheets, $\lambda$ represents the mean free path for inter-sheet collisions, $C_v$ the volumetric specific heat of individual nanosheets due to phonon mediated heat conduction, $\varphi$ is the volume percentage of Graphene loading and θ is a temperature dependant inter-sheet collision term. The dynamic conductivity is the manifestation of the thermal transport due to collisions among the particles, an inevitable event caused by the Brownian disturbance of the fluid molecules. The Brownian motion velocity of the sheets has been computed using Stokes-Einstein's model as given in Eq. (1). Since, $\overline{U_B}$ is the mean Brownian velocity; hence, based on kinetic theory assumption, $\overline{U_B}=3U_B$. The mean free path for inter-particulate collisions of nanosheets has been assumed to be of the order of $10^{-6}$ m, a reasonably valid assumption for sheet sizes of the order of $10^{-9}$ m. The volumetric specific heat for individual nanosheets, $C_v$, has been determined utilizing the Debye approximation model for the phonon density of states per unit volume. It has been proposed that unlike bulk materials, nanosheets exhibit temperature dependant $C_v$ if the temperature is low compared to the Debye temperature. At nanofluid operating temperature ranges, phonon transport is the dominant agent for in-plane heat conduction in Graphene. Under the Debye approximation, the volumetric specific heat[25], can be expressed as

$$C_v = 9k_B \left(\frac{N}{V}\right)\left(\frac{T}{\theta_D}\right)^3 \int_0^{\frac{\theta_D}{T}} \left(\frac{x^4 e^x dx}{(e^x-1)^2}\right) \qquad (8)$$

In Eq. (8), $N/V$ represents the number of atoms per unit volume, while $\theta_D$ represents the Debye temperature for Graphene for planar modes of phonon transport. At low temperatures, the upper limit for the integral in Eq. (8) can be assumed to approach infinity, reducing the equation to the form

$$C_v = \frac{36\pi^4 k_B}{15}\left(\frac{N}{V}\right)\left(\frac{T}{\theta_D}\right)^3 \qquad (9)$$

Since the value of $\theta_D$ for planar modes in Graphene is around 2300 K[26], and the operating range for most liquids are much lower than this value, Eq. (9) can be used without suffering appreciable errors. It may be also noted that $k_{dynamic}$, being a function of Brownian velocity, reduces rapidly with increasing sheet sizes.

The variable θ has been theorized to be a collision cross-section and denotes the number of effective elastic inter-sheet collisions occurring at any given instant of time, at the specified temperature, within the analytical domain. It is proposed to be a linear function of temperature, assuming the form of $\theta = aT - b$, where 'a' and 'b' are constants, whose values depend upon the properties of the base fluid and the dispersed media. It has been proposed that for every fluid – dispersed media pair, there exists a critical temperature, above which the inter-particulate collisions among the nanosheets can be considered elastic and θ is positive. Above this temperature, the present model is found to be significantly accurate. Below this critical temperature, the value of θ becomes negative, and $k_{dynamic}$ can no longer be incorporated while determining $k_{sd}$. It has been hypothesized that below the critical temperature, the collisions become inelastic and cause sheet agglomerations, leading to increase in effective sheet face size and lowering of $k_{EMT}$. This eventually reduces the value of $k_{sd}$, but it still remains greater than $k_{medium}$.

Validation with the experimental data of water-GNF[11] and with that of Ethylene Glycol(EG) based GNF[12] yields good results. The constants 'a' and 'b' are found to be consistent for all temperatures and volume concentration of Graphene loading when assigned values of 50 K$^{-1}$ and 15100 for Graphene in water. The values are constant for water-Graphene pair and are not adjustable. From the DLS studies[11], a weighted average based on the distribution pattern of sheets smaller than $L_{cr}$ provides an average sheet face size of 25 nm. For the EG-GNF, the specified sheet sizes[12] have been used for calculations. Tri-layer Graphene with the standard 0.335 nm inter-layer distance[1] has been considered. For such a nanosheet, the number of atoms N has been assumed to be of the order of $10^6$. These values have been utilized to predict $k_{sd}$.

The effective thermal conductivity of the GNF, $k_{gnf}$, is proposed to be the geometric mean of $k_{perc}$ and $k_{sd}$, and is expressed as

$$k_{gnf} = k_{sd}^{\alpha} \, k_{effective}^{(1-\alpha)} \qquad (10)$$

Since nanosheets suspended amongst fluid molecules can be considered to constitute a statistical population, the geometric mean provides a more normalized value of the thermal conductivity of the system than a simple arithmetic mean. From DLS analysis reported in[11], α is found to be 0.34. However, the work reported[12] contains no details of the sheet size distribution. A computer program is used to generate randomized distribution fractions (α) and based on a million such random values; an average value for $k_{sd}$ was deduced. Comparisons between the experimental data and the predicted enhancements based on the present model have been presented in Figs. (2) and (3). The temperature independence due to loss of dynamicity, resulting from the use of Graphene sample with micron sized sheets and due to the high viscosity of EG is clearly observed in Fig.(3).

**[FIGURE 2]**

**[FIGURE 3]**

The trends in the enhancement of thermal conductivity with varying 'α', 'φ' and temperature are presented in Figs. (4) and (5). Analysis of Fig.(5) reveals interesting shift in behavior of the GNF as 'α' reaches asymptotic limits of unity or zero. In the former configuration, percolation is the sole governing factor and the thermal conductivity response to temperature dies out, similar to CNT nanofluids, and can been seen in Fig.(5). In the latter case, particle dynamics is the sole player and the thermal conductivity exhibits sharp temperature response, similar to metallic or oxide nanofluids. However, as the percolating length of Graphene flakes is small, enhancement (at low concentrations) due to percolation is very low for GNFs.

**[FIGURE 4]**

**[FIGURE 5]**

In summary, the physical mechanisms behind heat conduction in GNFs have been explored and an analytical model, considering the poly-dispersed nature of GNFs, has been proposed to predict its thermal conductivity. It has been established that the thermal conductivity enhancement in GNF is due to the dual behavior of sheet percolation and Brownian motion assisted inter-sheet interactions. Graphene, being in the form of flakes exhibit behavioral duality, in between the likes of CNT (percolation) and metallic or metal oxide particles (particle dynamicity dominated). The critical sheet size that governs the inclination of the Graphene sample towards either phenomenon can be determined from the Stokes-Einstein's formula for Brownian diffusion and the settling velocity of particles in a fluid medium. The proposed dynamic conductivity is governed by the phonon mediated specific heat in Graphene and temperature dependant inter-sheet collisions. The model is found to accurately predict the thermal conductivity enhancement in GNFs. Plots evolved from the model (Figs.(4) and (5)) can predict the relationship between '$\alpha$', '$\theta$' and temperature, providing insight into the behavioral aspects of the GNFs.

To infer, as analysis suggests, Graphene samples with majority of the flakes in the order of 25 nm or lesser, when used in minute concentrations produce similar levels of enhancement that are obtained from larger concentrations of Graphene with appreciable percentage of flakes in the micron size range. Studies into methods to prepare Graphene samples consisting of only nanometer sheet sizes would result in manufacturing of 'smart' GNFs for enhanced thermal, electrical and possible targeted drug delivery systems at economical loading concentrations.

# References

*Electronic address: skdas@iitm.ac.in

[1] A.K.Geim and K.S.Novoselov, Nat. Mater. **6**, 183 (2007).
[2] T. T. Baby and S. Ramaprabhu, J. Appl. Phys. **108**, 124308 (2010).
[3] H.E.Patel; T.Sundararajan and S.K.Das, J. Nanopart. Res. **12**, 1015 (2010).
[4] J.A.Eastman et. al., Appl. Phys. Lett. **78** (6), 718 (2001).
[5] S. Lee et. al., J. Heat Transfer **121**, 280 (1999).
[6] X. Wang; X. Xu and S. U. S. Choi, J. Thermophys. Heat Transfer, **13**, 474 (1999).
[7] S. U. S. Choi, ASME Fluid Eng. **231**, 99 (1993).
[8] W. Yu; H. Hie and W. Chen, J. Appl. Phys. **107**, 094317 (2010).
[9] W. Yu; H. Hie and D. Bao, Nanotechnology **21**, 055705 (2010).
[10] P.Kim et al, Phys. Rev. Lett. **87**, 215502 (2001).
[11] Sen Gupta et. al., J. Appl. Phys. **110**, 084302 (2011).
[12] Wei Yu et. al., Phys. Lett. A. **375**, 1323 (2011).
[13] A. A. Balandin et al, Nano Lett. **8**, 902 (2008).
[14] S. K. Das et. al., J. Heat Transfer. **125**, 567 (2003).
[15] N N Venkata Sastry et. al., Nanotechnology **19**, 055704 (2008).
[16] P. Keblinski et. al., Int. J. Heat Mass Transfer. **45**, 855 (2002).
[17] R.Prasher et. al., Appl. Phys. Lett. **89**, 143119 (2006).
[18] S.P.Jang and S.U.S Choi, Appl. Phys. Lett. **84**, 4316 (2004).
[19] R.Prasher; P.Bhattacharya and P.E.Phelan, Phys. Rev . Lett. **94**, 025901 (2005).
[20] D. Hemanth Kumar et. al, Phys. Rev. Lett. **93**, 144301 (2004).
[21] C. Kleinstreuer and Y. Feng, Nanoscale Res. Lett. **6**, 229 (2011).
[22] H.E. Patel et. al., Pramana J. Phys. **65**, 863 (2005).
[23] Q. Xue, Phys. Lett. A. **307**, 313 (2003).
[24] Scott T. Huxtable et. al., Nat. Mater. **2**, 731 (2003).
[25] Gang Chen, Nanoscale Energy Transpot and Conversion : A parallel Treatment of Electrons , Molecules, Phonons and Photons.( Oxford University Press Inc., New York, 2005), p.145.
[26] V.K.Tewary and B.Yang, Phys. Rev. B. **79**, 125416 (2009).

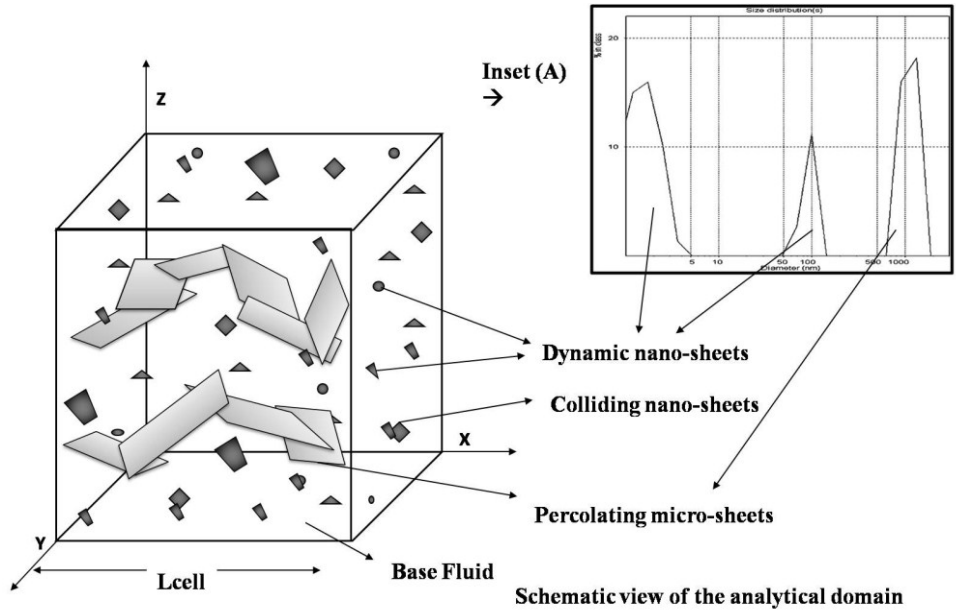

**FIG. 1.** The analytical domain
**Inset (A)** : DLS data. Sengupta. et. al. (2011)

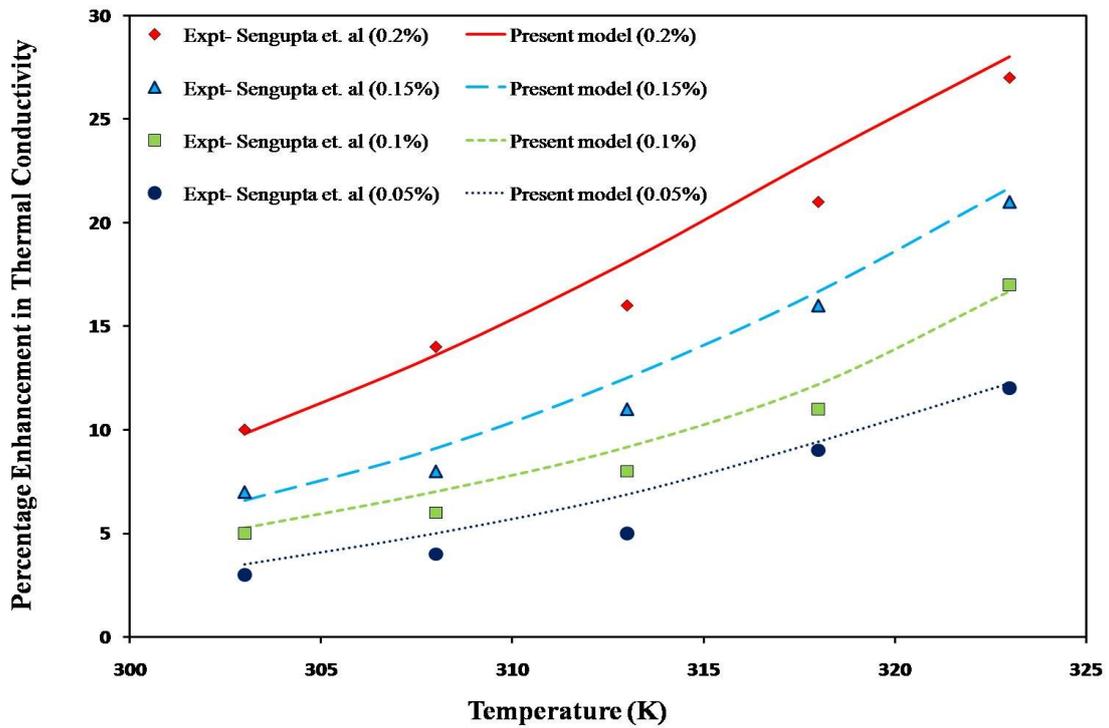

**FIG. 2.** Validation with Experimental results for water-GNF
($\varphi$ = 0.2%, 0.15%, 0.1% and 0.05%)

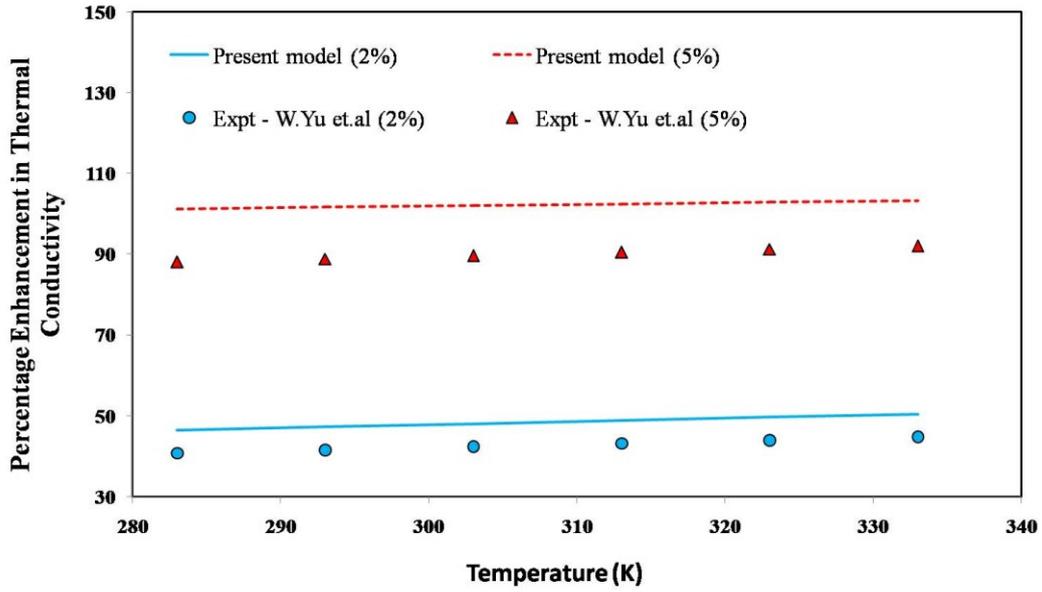

**FIG. 3.** Validation with Experimental results for EG (k = 0.25 W/mK) based GNF (φ = 2% and 5%)

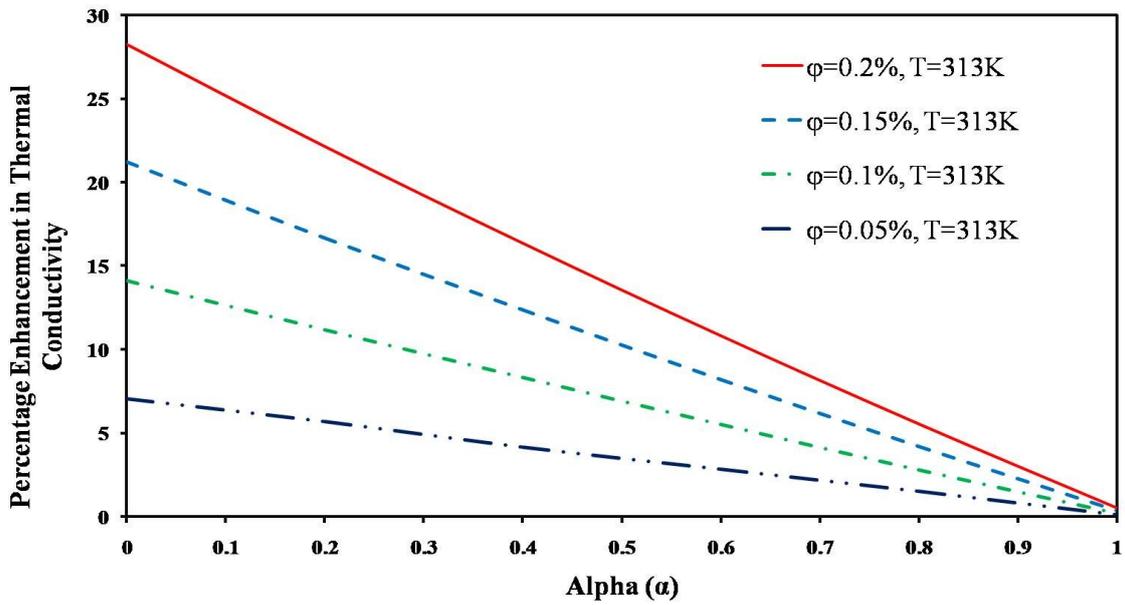

**FIG. 4.** Variation of Percentage enhancement of Thermal Conductivity with α and φ at constant Temperature

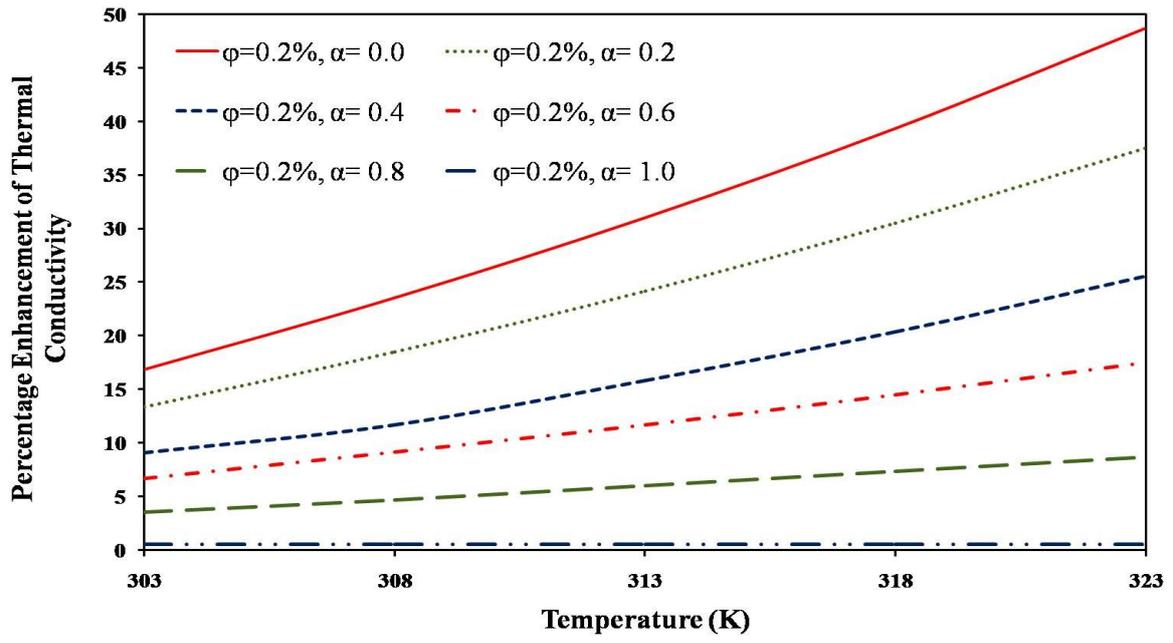

**FIG. 5.** Variation of Percentage enhancement of Thermal Conductivity with α and Temperature at constant φ